\documentclass{Interspeech2024}

\usepackage{amsmath,graphicx, paralist, xcolor, CJKutf8, enumitem}
\PassOptionsToPackage{hyphens}{url}\usepackage{hyperref}

\newcommand\figref[1]{Fig.~\ref{#1}}
\newcommand\tabref[1]{Table~\ref{#1}}
\newcommand\secref[1]{Section~\ref{#1}}
\newcommand{\mr}[1]{\mathrm{#1}}
\DeclareMathOperator*{\argmax}{arg\,max}

\interspeechcameraready

\title{Audio-conditioned phonemic and prosodic annotation \\for building text-to-speech models from unlabeled speech data}

\name[affiliation={1*}]{Yuma}{Shirahata}
\name[affiliation={1*}]{Byeongseon}{Park}
\name[affiliation={1}]{Ryuichi}{Yamamoto}
\name[affiliation={1}]{Kentaro}{Tachibana}

\address{
  $^1$LY Corporation., Japan
}
\email{yuma.shirahata@lycorp.co.jp}


\newcommand{\parkedit}[1]{\textcolor{black}{#1}}
\newcommand{\ppedit}[1]{\textcolor{black}{#1}}
\newcommand{\pppedit}[1]{\textcolor{black}{#1}}

\newcommand{\sedit}[1]{\textcolor{black}{#1}}
\newcommand{\ssedit}[1]{\textcolor{black}{#1}}
\newcommand{\sssedit}[1]{\textcolor{black}{#1}}
\newcommand{\redit}[1]{\textcolor{black}{#1}}

\begin{document}

\maketitle

\fontsize{9.0}{10.0}\selectfont
\renewcommand{\thefootnote}{\fnsymbol{footnote}}
\footnotetext[1]{Equal contribution.}
\renewcommand{\thefootnote}{\arabic{footnote}}
\begin{abstract}
This paper proposes an audio-conditioned phonemic and prosodic annotation model for building text-to-speech (TTS) datasets from unlabeled speech samples.
For creating a TTS dataset that consists of label-speech paired data, the proposed annotation model leverages an automatic speech recognition (ASR) model to obtain phonemic and prosodic labels from unlabeled speech samples.
By fine-tuning a large-scale pre-trained ASR model, we can construct the annotation model using a limited amount of label-speech paired data within an existing TTS dataset.
To alleviate the shortage of label-speech paired data for training the annotation model, we generate pseudo label-speech paired data using text-only corpora and an auxiliary TTS model. This TTS model is also trained with the existing TTS dataset.
Experimental results show that the TTS model trained with the dataset created by the proposed annotation method can synthesize speech as naturally as the one trained with a fully-labeled dataset.
\end{abstract}

\section{Introduction}
The field of Text-to-speech (TTS) has experienced significant progress owing to the rapid advancements of deep neural network-based approaches~\cite{tan2021survey}.

For training TTS models, a sufficient amount of speech-text paired data is essential. While collecting a large amount of unlabeled speech data is comparatively straightforward as demonstrated by the datasets used for training audio self-supervised learning (SSL) models~\cite{baevski2020wav2vec,hsu2021hubert,mohamed2022self}, the latter often necessitates accurate phonemic and prosodic labels for the development of high-quality TTS systems~\cite{le2024voicebox,yasuda2019investigation,pan2020unified}, which are challenging to obtain in large quantities. Thus, the acquisition of reliable labels from speech is crucial to leverage the vast amounts of unlabeled speech data in the TTS field.

To obtain phonemic and prosodic labels from unlabeled speech, a typical approach is the sequential application of automatic speech recognition (ASR) models followed by text processing~\cite{le2024voicebox,wang2023neural,zhang2023speak}: 1) employing ASR models that output grapheme sequences given unlabeled speech samples; 2) performing text-based processing such as grapheme-to-phoneme (G2P) conversion~\cite{chen2003conditional,chae2018convolutional} and prosody prediction~\cite{rosenberg2010autobi,park2022unified} on the output of the ASR model. A key advantage of this approach is the use of extensive dictionary data and ASR models trained on large text corpora. Nonetheless, the task of predicting phonemic and prosodic labels from grapheme sequences inherently presents a one-to-many mapping challenge, making accurate annotation difficult without audio information. This is because a text can be interpreted and vocalized in multiple ways, influenced by factors such as the speaker's dialect, age, \pppedit{and} speech disfluencies, among others.

On the other hand, there are some studies that utilize audio information to annotate prosodic \ssedit{labels} on speech samples for \ssedit{creating TTS datasets}~\cite{dai2022automatic,yuan2022low}. \sedit{These studies successfully improved the accuracy of prosody prediction owing to the information derived from input speech}. However, they are limited to scenarios where the correct text and phonemic information are provided. Research has not yet advanced to address performance on entirely unlabeled speech data, which represents a more realistic scenario.

To address the limitations of the previous works, this paper proposes an annotation model that predicts phonemic and prosodic labels (hereinafter \textit{TTS labels}) simultaneously from unlabeled speech data, conditioned on input speech information. 
\sedit{For creating a TTS dataset from unlabeled speech samples, the proposed annotation model leverages an ASR model to obtain TTS labels corresponding to the input speech samples. Specifically, we can construct the annotation model by fine-tuning a large-scale pre-trained ASR model with a limited amount of labeled speech data within an existing TTS dataset.
Furthermore, to address the challenge of amassing a sufficient amount of label-speech paired data for training the annotation model, we propose a data augmentation method utilizing TTS. In this method, an auxiliary TTS model is first trained on a limited amount of label-speech paired data within the existing TTS dataset, and the model is then used to generate pseudo label-speech paired data from text-only corpora. The combination of the pre-trained ASR model and data augmentation enables the construction of a model capable of generating highly accurate TTS labels, even with a limited amount of label-speech paired dataset.
For the architecture of the annotation model, we adopted the Transformer for its superior ability in sequence-to-sequence problems~\cite{vaswani2017attention}. The model receives raw speech sequences as input and predicts the corresponding TTS labels in an auto-regressive manner.
Once the annotation model is trained, it is applied to unlabeled speech samples to get the label-speech paired data for TTS model training. }

Through experiments, we find that the proposed method is able to annotate unlabeled speech more accurately than the baseline method that cascades an ASR model and text processing even when the number of the ground truth labels is less than 5,000 samples of \redit{a} single speaker (\parkedit{character error rate (}CER\parkedit{)} \pppedit{on phonemic label prediction}: 6.45\% vs. 2.44\%, \parkedit{$F_1$ score on p}rosodic label \ssedit{prediction: 68.51\% vs. 95.96\%)}. \parkedit{Moreover}, TTS models trained with the TTS datasets generated by the proposed method achieved comparable performance to those trained with the fully-labeled ƒdataset in terms of naturalness. \ssedit{Audio samples are available on our demo page\footnote{\sssedit{\url{https://yshira116.github.io/pp_annotation/}}}.}

\section{Method}\label{sec: proposed}
\subsection{Problem formulation} \label{subsec: formulation}
To train a TTS model from unlabeled speech data, this study aims to \ssedit{construct an annotation model that can} estimate a TTS label sequence $\bm{y} = \{y_m \in \mathcal{Y}\}_{m=1}^{M}$ from an unlabeled speech sample $\bm{X} = \{x_n \in \mathbb{R}^{D_{in}}\}_{n=1}^{N}$. Here, $\mathcal{Y}$ and ${M}$ are the vocabulary of TTS input tokens (i.e., a mixed vocabulary of phonemic and prosodic \parkedit{labels}) and the length of output \sedit{TTS labels}, $D_{in}$ and $N$ are the dimension\pppedit{s} of acoustic features of input speech and its length, respectively. In mathematical terms, we optimize the following conditional likelihood objective:
\begin{equation}
    L = p(\bm{y}|\bm{X}). \label{eq:yx}
\end{equation}

\begin{figure}[!t]
\begin{minipage}[b]{1.0\linewidth}
    \centering
    \includegraphics[width=0.72\linewidth]{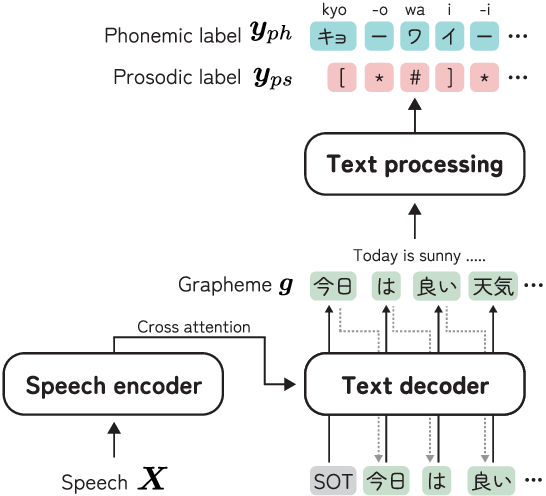}
        \centerline{(a) baseline framework}
        \medskip
        \vspace{-1mm}
\end{minipage}
\begin{minipage}[b]{1.0\linewidth}
    \centering
    \includegraphics[width=0.72\linewidth]{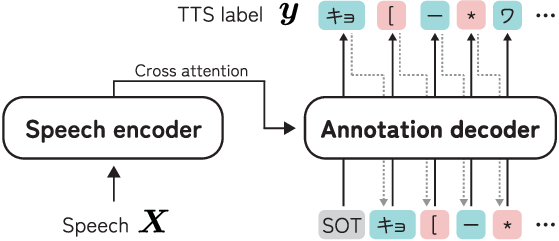}
    \centerline{(b) proposed framework}
    \medskip
    \vspace{-3mm}
\end{minipage}
\vspace{-5mm}
\caption{
    Overview of the baseline and the proposed annotation model. In the baseline framework, phonemic and prosodic labels are predicted from the grapheme sequence. \pppedit{In contrast}, they are predicted directly from speech in the proposed method.
}
\label{fig:overview}
\vspace{-2mm}
\end{figure}

However, since $\bm{y}$ is a mixed representation of multiple sequences and difficult to predict at once, the following conditional dependency assumption is typically introduced in previous works:
\begin{align}
    p(\bm{y}|\bm{X}) &= p(\bm{y}_{\mr{ph}}, \bm{y}_{\mr{ps}}|\bm{X}) \notag \\
    &= p(\bm{y}_{\mr{ph}}, \bm{y}_{\mr{ps}}|\bm{g})p(\bm{g}|\bm{X}), \label{eq:yx_sub}
\end{align}
where $\bm{g}$, $\bm{y}_{\mr{ph}}$, and $\bm{y}_{\mr{ps}}$ are the corresponding grapheme sequence, phonemic label sequence, and prosodic label sequence, respectively. In \eqref{eq:yx_sub}, since the first term is independent \pppedit{of} speech $\bm{X}$, it can be optimized using only text-based methods. In addition, since many high-quality grapheme-based ASR models are readily available online~\cite{radford2023robust,watanabe2018espnet}, the optimization of the second term is also straightforward. However, since the first term cannot consider the speech information to estimate the label sequence, this method is inherently accompanied by errors in G2P and prosodic label estimation, which results in a sub-optimal prediction. The overview of this method is depicted in \figref{fig:overview} (a). To overcome this problem, we propose a model that directly optimizes \eqref{eq:yx} in \ref{subsec: ant_model}.

\subsection{Annotation model} \label{subsec: ant_model}

The overview of the proposed annotation model is shown in \figref{fig:overview} (b). Following successful prior works that predict a mixture of multiple sequences as a single sequence~\cite{omachi2021end,audhkhasi2018building,shafey19_interspeech},
we adopted the encoder-decoder Transformer architecture as the base structure of the annotation model. 
The model is composed of the speech encoder and the annotation decoder. The speech encoder encodes the input acoustic feature sequence $\bm{X}$ into a hidden speech embedding sequence. The annotation decoder then generates the corresponding TTS label sequence $\bm{y}$ conditioned on the embedding sequence in an auto-regressive manner:
\begin{equation}
    \log p(\bm{y}|\bm{X}) = \sum_{m=1}^{M} \log p(y_m|y_1,\hdots,y_{m-1}, \bm{X}).
\end{equation}
The annotation model is trained on a paired dataset of $(\bm{X}, \bm{y})$\sedit{, to minimize the cross entropy loss of the model outputs and ground truth labels}. During inference, given an unlabeled speech sample $\bm{X}$, the model infers the corresponding \ssedit{TTS label} sequence $\bm{\hat{y}}$ as follows:
\begin{equation}
    \bm{\hat{y}} = \argmax_{y \in \mathcal{Y}^{*}} p(\bm{y}|\bm{X}),
\end{equation}
where $\mathcal{Y}^{*}$ denotes a set of all possible hypotheses. 

\subsection{Text-to-speech data augmentation} \label{subsec: ttsaug}

\begin{figure}[!t]
\begin{minipage}[b]{1.0\linewidth}
    \centering
    \includegraphics[width=0.45\linewidth]{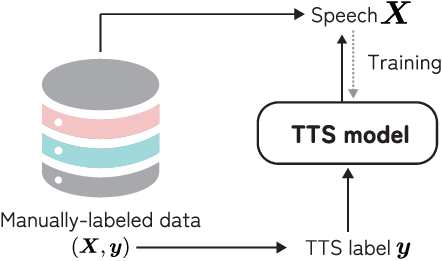}
        \centerline{(1) \ssedit{Auxiliary} TTS model training}
        \medskip
        \vspace{-2.5mm}
\end{minipage}
\begin{minipage}[b]{1.0\linewidth}
    \centering
    \includegraphics[width=0.68\linewidth]{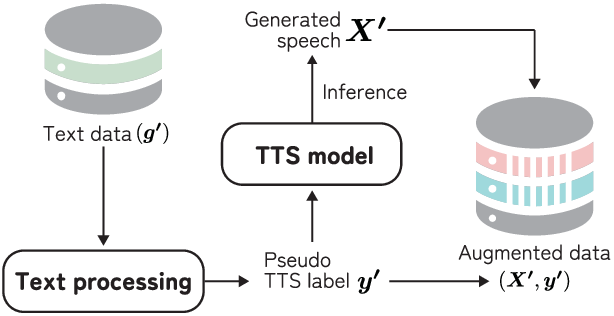}
    \centerline{(2) Data augmentation by \ssedit{the auxiliary} TTS model}
    \medskip
    \vspace{-3mm}
\end{minipage}
\vspace{-5mm}
\caption{
    Overview of the data augmentation method with the \ssedit{auxiliary} TTS model. (1) First, the \ssedit{auxiliary} TTS model is trained using \parkedit{manually }labeled data. (2) Then, the \ssedit{auxiliary} TTS model is used to generate augmented paired data from text\pppedit{-}only corpora.
}
\label{fig:ttsaug}
\vspace{-2mm}
\end{figure}

Although we can train the annotation model with paired data consisting of $(\bm{X}, \bm{y})$, amassing substantial annotated data often proves challenging. This is because accurately labeling speech samples requires specialized expertise and is notably time-consuming. To deal with this issue, we propose a data augmentation method using an \ssedit{auxiliary} TTS model. The overview of the proposed TTS data augmentation method is described in \figref{fig:ttsaug}. As shown in \figref{fig:ttsaug}, we first train an \ssedit{auxiliary} TTS model $\bm{M}$ with a limited size of label-speech paired dataset $\bm{D} = \{\bm{X_i}, \bm{y_i}\}_{i=1}^{K}$, where $K$ denotes the number of training samples with manually-annotated labels. Second, we prepare a large-scale text-only dataset $\bm{D'_g} = \{\bm{g'_i}\}_{i=1}^{K'}$ that has only grapheme sequences. Here, $\bm{K'}$ is the number of samples in the text dataset. Third, a text processing module is used to generate pseudo TTS labels $\bm{D'_y} = \{\bm{y'_i}\}_{i=1}^{K'}$ from $\bm{D'_g}$. Note that the text processing module here is not required to be correct, since the \ssedit{auxiliary} TTS model $\bm{M}$ is expected to generate speech samples that are faithful to input TTS labels. In other words, if the text processing module generates an incorrect phoneme sequence, the generated speech sample from it reflects the incorrect sequence, which is consistent as paired data for \ssedit{the training of the annotation model}. Finally, $\bm{M}$ generates augmented speech samples $\{\bm{X'_i}\}_{i=1}^{K'}$ from $\bm{D'_y}$, and augmented training data $\bm{D'} = \{\bm{X'_i}, \bm{y'_i}\}_{i=1}^{K'}$ is obtained.

\section{Experiments}\label{sec:exp}
To assess the performance of the proposed methods, we conducted two types of experiments. \secref{subsec: unlabel_annt} objectively evaluates the accuracy of TTS labels generated from unlabeled speech datasets. \secref{subsec: exp_tts} investigates the performance of the proposed method when applied to TTS tasks. 
\subsection{Annotation of unlabeled speech data}\label{subsec: unlabel_annt}
\subsubsection{Experimental conditions}\label{subsubsec: exp_cond}
\textbf{Dataset and pre-processing:}
For the training of the proposed annotation models, two datasets were prepared \sedit{to investigate the performance of the models when trained on 1) a limited amount of labeled data, and 2) a large scale data with a variety of speakers}. \sedit{For the former, we adopted} JSUT, which is a public Japanese speech corpus uttered by a single female speaker~\cite{sonobe2017jsut}. We used the {\it basic5000} subset and its manual \parkedit{TTS} labels\footnote{\url{https://github.com/sarulab-speech/jsut-label}}. The dataset consists of 5,000 text samples and 6.78 hours of speech. We split the data into 4,500 and 250 samples for training and validation, respectively. The remaining 250 samples were not used in this experiment. \sedit{For the latter, we used} proprietary Japanese speech corpora recorded by six male and eleven female Japanese professional speakers with manually annotated labels\pppedit{. The corpora} consist of \sedit{173,987} samples and \sedit{207.96} hours of speech. We held out the samples of two male\pppedit{s} and two female\pppedit{s} for evaluation\pppedit{,} and the other speakers' data was used for training and validation. The number of data for training, validation, and evaluation were 153,551, 4,449, and 15,987, respectively. \sedit{Hereinafter, this dataset will be referred to as LARGE.}

\textbf{TTS data augmentation:}
In our experiment, TTS data augmentation was applied to the JSUT dataset. 
The model architecture of the TTS model \sedit{for data augmentation} was based on Period VITS~\cite{shirahata2023period}. We used the same configuration that will be described in \ref{subsubsec: exp_cond2}. To exclude the bias of the text domain, the augmented text data $\bm{D'_g}$ was taken from the training set of the LARGE dataset (153,551 samples). For the text processing module, Open JTalk\footnote{\url{https://open-jtalk.sp.nitech.ac.jp/}} was used. The total amount of augmented speech data was 115.5 hours. \sssedit{Note that the augmented samples by the TTS model trained on the JSUT dataset generally had a faster speed than the LARGE dataset, which resulted in a smaller data size for the same text set.}

\begin{figure}[t]
\centering
\includegraphics[width=0.85\linewidth]{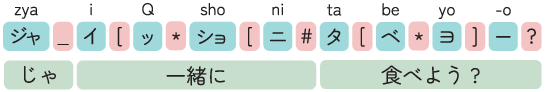}
\begin{CJK}{UTF8}{min}
\caption{
    Example of TTS labels \sedit{for a Japanese text} ``じゃ一緒に食べよう?'' (Well, let's eat together). The blue, red, and green squares \parkedit{denote} phonemic labels, prosodic labels, and grapheme, respectively.
}
\label{fig:prosody_label}
\end{CJK}
\end{figure}

\textbf{phonemic/prosodic labels:}
For phonemic \pppedit{and} prosodic \pppedit{labels}, we used Kurihara et al.~\cite{kurihara2021}'s \pppedit{design}, as depicted in \figref{fig:prosody_label}.
In \pppedit{the} method, the prosodic status of each mora is represented by five labels considering the rules of the Japanese pitch accent in the Tokyo dialect.
The details of the labels are as follows:
\begin{inparaenum}[(1)]
    \item Pause (``\_'' in \figref{fig:prosody_label});
    \item Low to high accent change (``$[$'' in \figref{fig:prosody_label});
    \item High to low accent change (``$]$'' in \figref{fig:prosody_label});
    \item Accentual phrase boundary (``\#'' in \figref{fig:prosody_label});
    \item Raise-type boundary pitch movement (for question sentence, ``$?$'' in \figref{fig:prosody_label}).
\end{inparaenum}
In this experiment, we additionally introduced a padding token for a mora that does not apply to the five categories above (``*'' in \figref{fig:prosody_label}).
\parkedit{For the phonemic labels, we used Japanese katakana characters to represent the Japanese phonemic status of each mora.}

\textbf{Model details:}
All the proposed annotation models were fine-tuned from the encoder-decoder-based public speech recognition model Whisper~\cite{radford2023robust}. We used the {\it small}\footnote{Larger models were not used due to the limitation of computational resources.} model for all the experiments.  We fine-tuned each model for 100k steps, with a batch size of 36. The learning rate was increased to 0.0002 with warm-up steps of 500, and then linearly decreased to reach zero at the 100k step. The parameters in the encoder part were frozen during fine-tuning to stabilize the training.  Model checkpoints with the best validation loss were used for the evaluation. In addition to the proposed models, two text-based baseline models were also prepared. The systems used in our experiments are summarized below:
\begin{description}
\item[ANNT-JSUT:] Proposed annotation model trained on the manually annotated JSUT training data.
\item[ANNT-JSUT-TTSAUG:] Proposed annotation model trained on the manually annotated JSUT training data and TTS augmentation data. 
\item[ANNT-LARGE:] Proposed annotation model trained on the manually annotated LARGE training data.
\item[ASR-NLP:] A baseline model that obtains grapheme transcription by Whisper small model and performs text-based post-processing to get TTS labels.
\item[GT-NLP:] A baseline model that obtains grapheme transcription from ground truth text data and performs text-based post-processing to get TTS labels.
\end{description}
For \textbf{ASR-NLP} and \textbf{GT-NLP}, Open JTalk was used to obtain the TTS labels from grapheme sequences.

\subsubsection{Evaluation on annotation accuracy}\label{subsubsec: eval_annotation}

To evaluate the performance of our proposed method on annotation tasks, we tested the models with 15,987 \ssedit{speech} samples in our dataset.
We used CER and $F_1$ scores as metrics to evaluate the phonemic and prosodic label annotation tasks, respectively.
To independently evaluate phonemic and prosodic label annotation tasks, we separated phonemic and prosodic information from \sssedit{manually annotated} ground truth and predicted labels.
Hence, we only used phonemic labels for the calculation of CER.
\pppedit{Since} it is impossible to compare ground truth prosodic labels and predicted labels when predicted phonemic labels are \pppedit{corrupted}, 
we only used \ppedit{4,379} samples, in which all models correctly predicted phonemic labels \pppedit{of} the test set, to evaluate the prosodic label annotation task.
\pppedit{
Additionally, for a fair comparison of the proposed and baseline methods, we excluded two prosodic labels on the evaluation:
\begin{inparaenum}[(1)]
    \item Pause;
    \item Raise-type boundary pitch movement.
\end{inparaenum}
This is because ground truth texts of \textbf{GT-NLP} include these labels as punctuation.
}

Table~\ref{tab:objective_eval_result} shows the performance of the models on \pppedit{TTS} label annotation tasks.
The findings are summarized as follows:

\begin{description}[style=unboxed,leftmargin=0cm]
    \item[Baseline vs. Proposed model]
    \pppedit{
    As shown in Table~\ref{tab:objective_eval_result}, the proposed model performed best in both metrics when a large amount of annotation data is available (i.e., \textbf{ANNT-LARGE}).
    Furthermore, all our proposed models outperformed the baseline methods on the prosodic label prediction tasks, even when the ground truth grapheme sequence is used in the latter (i.e., \textbf{GT-NLP}).
    The results imply that the utilization of audio information is significantly effective in TTS label prediction.
    }

    \item[Effectiveness of data augmentation]
    Table~\ref{tab:objective_eval_result} also shows that the proposed model trained with the augmented data by our framework (i.e., \textbf{ANNT-JSUT-TTSAUG}) significantly outperformed the baseline methods and the model trained with limited-scale data (i.e., \textbf{ANNT-JSUT}).
    This confirms that the proposed TTS data augmentation method improves the performance of the annotation model, \sedit{even if the augmented data is automatically generated from text-only corpora.}
\end{description}

\begin{table}[t]
    \centering
    \caption{Objective evaluation results on each task. CER and \ssedit{Prosody} $F_1$ are metrics for phonemic and prosodic label annotation tasks, respectively.}
    \scalebox{0.83}{
        \begin{tabular}{@{}lcc@{}}
        \toprule
        \textbf{Model}	& \textbf{CER ($\downarrow$)} &  \textbf{\ssedit{Prosody} $\mathbf{F_1}$ ($\uparrow$)} \\
        \midrule
        \textbf{ASR-NLP} & 6.45\% & \pppedit{68.51\%} \\
        \textbf{GT-NLP} & 2.53\% & \pppedit{73.43\%} \\
        \midrule
        \textbf{ANNT-JSUT} & 6.12\% & \pppedit{88.77\%} \\
        \textbf{ANNT-JSUT-TTSAUG} & \textbf{2.44\%} & \textbf{\pppedit{95.96\%}} \\
        \midrule
        \textbf{ANNT-LARGE} & \textbf{0.54\%} & \pppedit{\textbf{98.84\%}} \\
        \bottomrule
        \end{tabular}
    }
    \label{tab:objective_eval_result}
\end{table}
\vspace{-4mm}

\subsection{Application to text-to-speech} \label{subsec: exp_tts}

\subsubsection{Experimental conditions}\label{subsubsec: exp_cond2}
\sedit{To investigate the robustness of the proposed method against dataset variation, }three datasets were used for TTS experiments: JSUT, JVS ~\cite{takamichi2019jvs}, and the LARGE dataset described in \ref{subsubsec: exp_cond}. For JSUT, we split the data into 4,500, 250, and 250 samples for training, validation, and evaluation, respectively. Note that \textbf{ANNT-JSUT} and \textbf{ANNT-JSUT-TTSAUG} were excluded from the evaluation on JSUT as these models used the same dataset for the training of annotation models. For JVS, we split the samples of \textit{parallel100} subset into 90 and 10 samples for each speaker for training and validation, respectively. For testing, \textit{nonpara30} subset was used. For 
LARGE, the held-out data in \ref{subsubsec: exp_cond} was used for TTS experiments. The 15,987 samples were split into 14,000, 1,000, and 987 samples for training, validation, and evaluation, respectively.

We adopted the Period VITS architecture for our TTS model due to its high-quality speech generation capability ~\cite{shirahata2023period}. We followed the settings of the original paper with two exceptions: 1) we did not use an emotion encoder, since no emotional dataset was used in the TTS experiments; 2) the training step was set to 200k based on the results of preliminary experiments. Since Period-VITS requires duration information of each phoneme, we trained a forced alignment model based on Gaussian mixture model and \parkedit{h}idden Markov model (GMM-HMM)\parkedit{~\cite{baum1972inequality}} on ReazonSpeech dataset\parkedit{~\cite{yin2023reazonspeech}}, and used it to obtain phoneme alignment.

In addition to the \ssedit{TTS labels} generated by the models in \ref{subsec: unlabel_annt}, two types of \ssedit{TTS labels} were used in TTS experiments:

\begin{description}
\item[ORACLE:] This model uses manually annotated labels.
\item[ORACLE-WO-ACC:] This model uses manually annotated labels, but drops \sssedit{prosodic} labels. This model was introduced to assess the importance of prosodic labels.
\end{description}
Since manual annotation data was unavailable for JVS dataset, \textbf{ORACLE} was not trained, and \textbf{ORACLE-WO-ACC} was substituted with the phoneme sequences from Open JTalk with ground truth text. This model is referred to as \textbf{GT-NLP-WO-ACC}.

\subsubsection{Evaluation on Text-to-speech}\label{subsubsec:eval_tts}

\begin{table}[t]
    \centering
    \caption{
    MOS test results on different datasets with 95\% confidence intervals. Note that \textbf{Reference} denotes recorded speech samples.
    }
    \scalebox{0.83}{
    \begin{tabular}{@{}lccc@{}}
    \toprule
    \textbf{Model}	& \textbf{JSUT} & \textbf{JVS} & \textbf{LARGE} \\
    \midrule
    \textbf{GT-NLP-WO-ACC}      & -                                             & $2.52 {\scriptstyle\pm{0.11}}$            & - \\
    \textbf{ORACLE-WO-ACC}      & $2.79 {\scriptstyle\pm{0.11}}$                & -                                         & $3.36 {\scriptstyle\pm{0.12}}$ \\
    \midrule
    \textbf{ASR-NLP}            & $3.65 {\scriptstyle\pm{0.11}}$                & $3.43 {\scriptstyle\pm{0.11}}$            & $4.04 {\scriptstyle\pm{0.09}}$ \\
    \textbf{GT-NLP}             & $3.69 {\scriptstyle\pm{0.10}}$                & $3.75 {\scriptstyle\pm{0.10}}$            & $4.05 {\scriptstyle\pm{0.09}}$ \\
    \midrule
    \textbf{ANNT-JSUT}          & -                                             & $3.77 {\scriptstyle\pm{0.10}}$            & $4.26 {\scriptstyle\pm{0.08}}$ \\
    \textbf{ANNT-JSUT-TTSAUG}   & -                                             & $\textbf{3.95} {\scriptstyle\pm{0.09}}$   & $\textbf{4.33} {\scriptstyle\pm{0.08}}$ \\
    \textbf{ANNT-LARGE}         & $\textbf{4.11} {\scriptstyle\pm{0.09}}$       & $3.75 {\scriptstyle\pm{0.10}}$            & $4.29 {\scriptstyle\pm{0.09}}$ \\
    \midrule
    \textbf{ORACLE}             & $4.15 {\scriptstyle\pm{0.09}}$                & -                                         & $4.22 {\scriptstyle\pm{0.09}}$ \\
    \midrule
    \textbf{Reference}          & $3.99 {\scriptstyle\pm{0.10}}$                & $4.39 {\scriptstyle\pm{0.09}}$            & $4.64 {\scriptstyle\pm{0.07}}$ \\
    \bottomrule
    \end{tabular}
    }
    \label{tab:mos_evaluation_result_all}
    \vspace{-3mm}
\end{table}

To evaluate the effectiveness of our proposed method on TTS tasks, we conducted subjective listening tests on the generated samples. These tests were based on the mean opinion score (MOS) of a five-point scale: 1 = Bad; 2 = Poor; 3 = Fair; 4 = Good; and 5 = Excellent. We asked native Japanese raters to make a quality judgment in terms of prosodic naturalness and pronunciation correctness. We showed the grapheme text to the raters during the listening tests to help accurately judge the naturalness of the prosody and pronunciation. The number of raters was eleven. 
For each of the three datasets, 50 sentences were randomly chosen from the evaluation set.
Then, ground truth labels were used\footnote{\sssedit{Using labels from the text processing model would be another option, but we used ground truth labels to minimize the errors derived from the input labels and focus on the quality of \redit{the} TTS models.}} to generate speech samples for each system. Since ground truth labels for JVS dataset were unavailable, we manually annotated the evaluation set.

\tabref{tab:mos_evaluation_result_all} summarizes the results of subjective evaluation.
\pppedit{Firstly, the MOS scores are significantly lower for the TTS models lacking accent information than the others. This confirms that prosodic labels are quite important in improving the naturalness of Japanese speech synthesis, as reported in previous works~\cite{yasuda2019investigation, park2022unified, kurihara2021}.}
We can also see that the proposed methods constantly outperform \redit{the} baseline methods on all datasets, which is consistent with the results of objective evaluation on annotation accuracy. Moreover, for JSUT and LARGE datasets, \redit{the} TTS models trained on the labels generated by proposed methods perform comparable or slightly better than those trained on oracle labels. This result indicates that the proposed method has the capability to generate a sufficiently high-fidelity TTS system from unlabeled speech data. Interestingly, for JVS dataset, \textbf{ANNT-JSUT-TTSAUG} achieved a higher score than \textbf{ANNT-LARGE}, which performed the best in objective evaluation.
\ssedit{One possible reason is that while the proposed TTS data augmentation method can generate consistent label-speech paired data through the auxiliary TTS model, manually annotated labels could be noisy due to the inconsistent annotation across multiple annotators, which made it difficult for the annotation model to learn the correct mapping. This result also suggests that the proposed TTS data augmentation method is still effective when applied to TTS tasks.}
For JSUT dataset, some TTS models got higher scores than the reference. This is likely due to the inclusion of unclear pronunciations and lip noise in some of the reference audio samples.

\section{Conclusions}
In this paper, we proposed an annotation model for building high-fidelity TTS systems from unlabeled speech data. The proposed model predicts phonemic and prosodic label sequences from speech input. To address the challenge of collecting a sufficient amount of labeled data for model training, a data augmentation method utilizing the TTS model was proposed. 
\sssedit{The proposed model generated accurate TTS labels, enabling high-quality TTS models even when the number of manually annotated data is limited.}
Future work includes applying our approach to more challenging speech samples, including those with emotional content or pronounced dialectal variations.

\bibliographystyle{IEEEtran}
\bibliography{refs}

\end{document}